\documentclass{Interspeech2024}

\interspeechcameraready

\title{Whisper-PMFA: Partial Multi-Scale Feature Aggregation for \\ Speaker Verification using Whisper Models}

\name[affiliation={1}]{Yiyang}{Zhao}
\name[affiliation={2}]{Shuai}{Wang}
\name[affiliation={3}]{Guangzhi}{Sun}
\name[affiliation={1}]{Zehua}{Chen}
\name[affiliation={1}]{Chao}{Zhang}
\name[affiliation={1}]{\\Mingxing}{Xu}
\name[affiliation={1*}]{Thomas Fang}{Zheng}

\address{
  $^1$Tsinghua University, China,
  $^2$Shenzhen Research Institute of Big Data, China \\
  $^3$University of Cambridge, United Kingdom\thanks{\hspace{-1ex}*Corresponding author}}
\email{ zhaoyy22@mails.tsinghua.edu.cn, wangshuai@cuhk.edu.cn, gs534@cam.ac.uk, \{zhc23,cz277,xumx,fzheng\}@tsinghua.edu.cn}


\usepackage{multirow}
\usepackage{cite}

\begin{document}

\maketitle


\begin{abstract}
In this paper, Whisper, a large-scale pre-trained model for automatic speech recognition, is proposed to apply to speaker verification. 
A partial multi-scale feature aggregation (PMFA) approach is proposed based on a subset of Whisper encoder blocks to derive highly discriminative speaker embeddings.
Experimental results demonstrate that using the middle to later blocks of the Whisper encoder keeps more speaker information. On the VoxCeleb1 and CN-Celeb1 datasets, our system achieves 1.42\% and 8.23\% equal error rates (EERs) respectively,
receiving 0.58\% and 1.81\% absolute EER reductions over the ECAPA-TDNN baseline, and 0.46\% and 0.97\% over the ResNet34 baseline. Furthermore, our results indicate that using Whisper models trained on multilingual data can effectively enhance the model's robustness across languages. Finally, the low-rank adaptation approach is evaluated, which reduces the trainable model parameters by approximately 45 times while only slightly increasing EER by 0.2\%.\footnote{Our source code will be released in Wespeaker.}

\end{abstract}

\section{Introduction}

Speaker verification (SV) is crucial for biometric authentication, aiming to confirm a person's identity based on their voice characteristics. In the past decade, the advent of deep learning has led to significant advancements in speaker verification technology\cite{snyder2018x,variani2014deep,dehak2010front}. Models such as the convolutional neural network-based Residual Network (ResNet) \cite{fang2021deep}, the time-delay neural network-based ECAPA-TDNN \cite{desplanques2020ecapa}, and their diverse variants\cite{zhou2021resnext,gu2023memory,tang2019deep}, alongside multi-scale feature fusion models like MFA-Conformer \cite{zhang2022mfa}, have significantly contributed to the development of the field. 
Meanwhile, novel training methods, including loss functions \cite{wang2018additive,deng2019arcface,liu2017sphereface}, strategic training approaches \cite{thienpondt2021idlab}, and score normalization techniques \cite{matejka2017analysis}, have also considerably enhanced the performance of speaker verification systems. 


However, the increasing complexity of model architectures intensifies the demand for training data\cite{hu2021model}. Acquiring this data, particularly labelled datasets, is far from trivial, this challenge is further exacerbated in the realm of low-resource languages. To address these challenges, researchers have proposed numerous solutions, a widely used strategy involves leveraging large pre-trained models trained on extensive corpora(e.g. Whisper\cite{radford2023robust}, HuBERT\cite{hsu2021hubert}, WavLM\cite{chen2022wavlm}). The integration of pre-trained models provides a robust foundation for feature extraction and representation learning, thereby alleviating some of the constraints imposed by data scarcity. Berns et al. implemented a speaker change detection task on Whisper and Wav2vec2 by innovatively adding speaker change labels to the training data\cite{berns2023speaker}. Further, Cai et al. explored the feasibility of applying automatic speech recognition (ASR) models to speaker verification tasks by testing an ASR-pretrained Conformer model in speaker verification scenarios\cite{cai2023pretraining,liao2023towards}. 

As a large pre-trained ASR model trained on extensive corpora, Whisper has been extensively trained on multilingual and diverse situational audio datasets\cite{radford2023robust}. This extensive training grants it impressive performance and robust cross-linguistic features. Accordingly, we adapted it as a pre-trained model for speaker recognition tasks.

However, due to the fine-tuning paradigm's intensive memory and computational resource requirements, leveraging large-scale models trained on extensive datasets introduces a significant challenge. Addressing this challenge, Sang et al. introduce an adapter approach for adapting self-supervised speech models to speaker verification tasks\cite{sang2024efficient}. Concurrently, Peng et al. explore parameter-efficient transfer learning methods for adapting pre-trained Transformer models to speaker verification tasks, aiming to reduce the computational burden\cite{peng2023parameter}.

In this paper, we propose an effective Whisper-based partial multi-scale feature aggregation model (Whisper-PMFA). Compared to the widely used MFA architecture \cite{zhang2022mfa}, our approach benefits from partial layer selection, effectively reducing the performance degradation caused by the integration of excessive irrelevant information. This selective process also mitigates the computational and memory overhead caused by full concatenation. Additionally, to address the issue of computational overhead, we explored the use of the low-rank adaptation (LoRA) approach\cite{hu2021lora} as an alternative to fine-tuning. The primary contributions of this paper can be summarized as follows:

\begin{itemize}
  \item We developed a partial multi-scale feature aggregation module, adapting the Whisper model for the speaker verification task through selective aggregation of Whisper layers.
    \item In our enhanced model, by incorporating LoRA\cite{hu2021lora} in place of comprehensive fine-tuning, we significantly reduced the model's trainable parameters by approximately 45 times while only incurring a marginal increase in EER of 0.2\%.
  \item Our experiments confirmed Whisper's potential for cross-lingual speaker recognition applications.
  \item Our evaluations on the VoxCeleb1 and CN-Celeb1 datasets have conclusively demonstrated that the proposed model achieves significant improvements over the baseline models.
\end{itemize}


\begin{figure*}[htbp]
  \centering
  \includegraphics[width=0.9\textwidth]{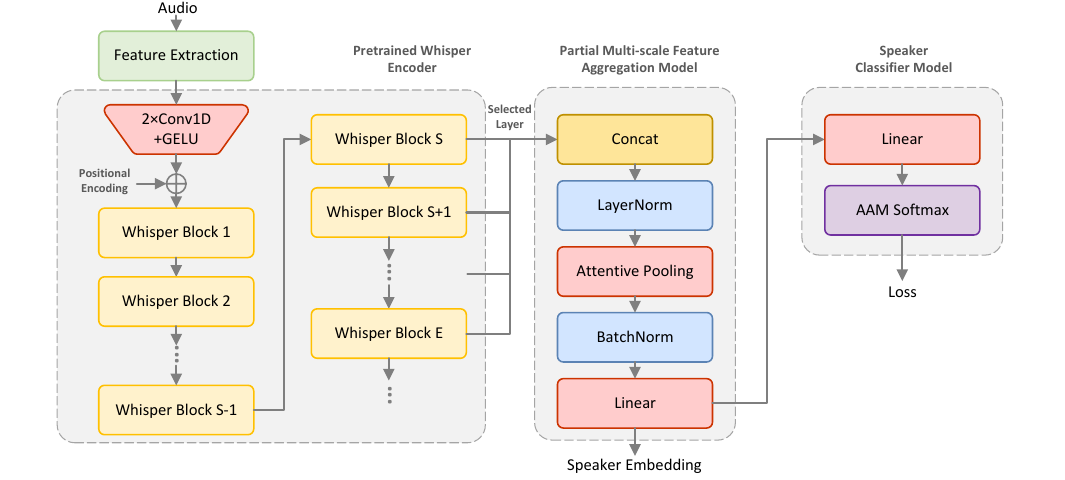}
  \caption{The overall architecture of Whisper-PMFA, where $S$ denotes the index of the initial Whisper block selected for feature aggregation, and $E$ represents the index of the final Whisper block selected.}
  \vspace{-1em}
  \label{fig:double-column-figure}
\end{figure*}

\section{Whisper-PMFA}

In this section, the Whisper-PMFA approach is introduced, aimed at using the rich and diverse speech knowledge obtained from a large amount of training data and embedded in a pre-trained ASR model to assist in the enhancement of speaker verification tasks. It combines the Whisper pre-trained model and aggregation model, with architectural details in Figure 1.

\subsection{Preliminary of Whisper}

Whisper\cite{radford2023robust} is a sophisticated, multilingual speech recognition model trained on a vast corpus of 680,000 hours of multilingual and diverse acoustic data. It combines a classical encoder-decoder Transformer\cite{vaswani2017attention} architecture, with its encoder comprising two convolutional layers, sinusoidal positional encoding, and a series of Transformer blocks, to effectively handle diverse linguistic and acoustic challenges. In this paper, we utilize the encoder from the Whisper Large-v2 as our pre-trained model, which comprises 32 Transformer blocks, each equipped with an attention mechanism that consists of 20 heads.

\subsection{Partial Multi-scale Feature Aggregation Model}

Multi-scale feature aggregation (MFA), involves the concatenation of output features from various frame-level modules within a speaker embedding architecture, before pooling at the utterance level. In previous research, MFA-Conformer has already demonstrated its effectiveness on Conformer models\cite{zhang2022mfa}. However, as the number of layers and the output size increase, concatenating the outputs of all blocks results in significant computational and memory overhead. Moreover, performing full concatenation may also introduce a substantial amount of non-speaker-related information, potentially leading to a degradation in model performance.

Therefore, unlike MFA, we replace the full concatenation operation with the partially selected layer concatenation:

\vspace{-5pt}
\begin{equation}
\begin{aligned}
    \mathbf{H}' &= \textrm{Concat}(h_{s},h_{s+1},\cdots ,h_{e}) \\
    \mathbf{H} &= \textrm{LayerNorm}(\mathbf{H}')
\end{aligned}
\end{equation}

\noindent where $s$ is the first Whisper block number to be selected, and $e$ is the last Whisper block number to be selected. $\mathbf{h}_{i}\in \mathbb{R}^{d\times T}$ is the output of $i$-th Whisper block, $\mathbf{H},\mathbf{H}'\in \mathbb{R}^{D\times T}$ with $D=k*d$, $k$ is the sum of the chosen Whipser block numbers.
Thereafter, we use attentive statistics pooling\cite{okabe2018attentive} to extract speaker cues from frame-level features that are helpful for the speaker verification task.
Finally, the speech vector is passed through batch normalization and a fully connected layer to obtain a low-dimensional speaker embedding representation.

\subsection{LoRA for model adaptation}


Compared to full fine-tuning, LoRA\cite{hu2021lora} optimizes storage and computational efficiency by modulating low-rank subspace parameters. In this paper, we apply LoRA to the Q (query), K (key), V (value), and O (output) weights of the Whisper model's multi-head attention, freezing the model's remaining parameters. For the model's weight matrix $W\in \mathbb{R}^{d\times k}$, the update mechanism is delineated as follows:

\vspace{-5pt}
\begin{equation}
\begin{aligned}
    W+\Delta W=W+BA
\end{aligned}
\end{equation}
\vspace{-1em}

\noindent where $B\in \mathbb{R}^{d\times r}$ is initialised with zeros, $A\in \mathbb{R}^{r\times k}$ is initialised with  random Gaussian initialisation, with the rank $r\ll \mathrm{min}(d,k)$.

\vspace{-1em}

\begin{figure}[htbp]
  \centering
  \includegraphics[width=0.45\textwidth]{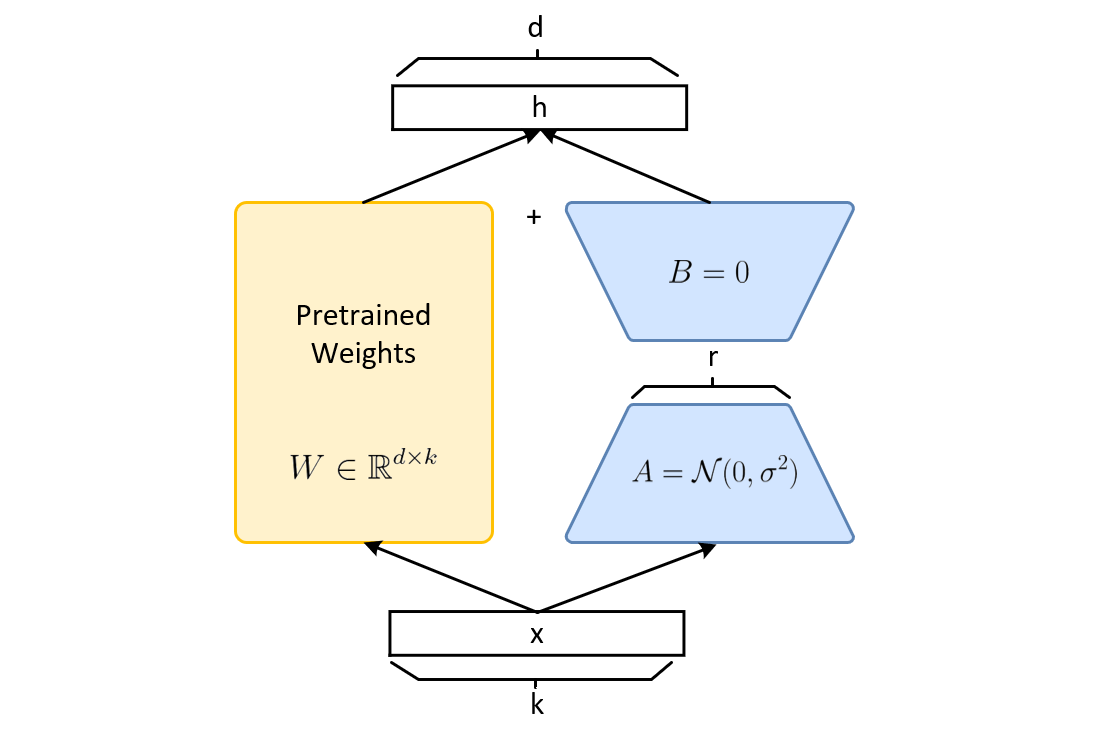}
  \caption{LoRA. The pretrained weight parameters $W$ are frozen, with only $A$ and $B$ being updated.}
  \label{fig:figure2}
\end{figure}

\vspace{-1.5em}





\section{Experimental setup}

\subsection{Dataset}
To investigate the effectiveness of the proposed methods, we conduct experiments on the VoxCeleb1\cite{nagrani2020voxceleb} and CN-Celeb1\cite{fan2020cn} datasets.
The VoxCeleb1 dataset is a large-scale audio-visual collection designed for speaker recognition tasks, containing over 150,000 utterances from 1,251 celebrities sourced from YouTube videos.
The CN-Celeb1 dataset is a comprehensive audio dataset tailored for speaker recognition research. It consists of approximately 130,000 utterances by 1,000 Chinese celebrities from various domains, including but not limited to entertainment, sports, and politics. 

During the training phase, we adhered to the official dataset partitions provided, training two distinct models on VoxCeleb1 and CN-Celeb1, respectively. In the testing phase for VoxCeleb1, we utilized the original trial list associated with VoxCeleb1 (VoxCeleb1-O) for evaluation. Similarly, for CN-Celeb1, we conducted tests using the official trial lists provided.

\subsection{Model configuration}

In our experiments, we selected two baseline models: ResNet34 and ECAPA-TDNN. Both baseline models, along with our proposed model, were implemented within the Wespeaker framework\cite{wang2023wespeaker}. The specifications and parameter settings for each model are as follows:

\noindent\textbf{ECAPA-TDNN}\cite{desplanques2020ecapa}: ECAPA-TDNN is a modified version of the Time Delay Neural Network (TDNN), it features an enhanced Channel-wise Correlation Matrix Attention mechanism and time-domain x-vectors. Our implementation adopted the recommended parameters from Wespeaker.

\noindent\textbf{ResNet34}\cite{zeinali2019but}: ResNet-based r-vector is the winning system of VoxSRC 2019. It utilizes residual connections to alleviate the vanishing gradient problem during training, enabling the effective training of very deep neural networks. The parameter settings also follow recipes in Wespeaker.

\noindent\textbf{Whisper-PMFA}: 
Our proposed model Whisper-PMFA is built upon the pretrained Whisper large-v2 model, which consists of 32 Transformer blocks. For our implementation, we selected partial blocks from the Whisper large-v2 model to obtain the speaker embedding. The details of the experiments and results will be presented in Section 4.2. The dimensionality of the speaker embedding was set to 192.

\begin{table*}[htbp]
\centering
\setlength{\tabcolsep}{5pt}
\caption{Intra-Language and Cross-Language Performance Evaluation for Systems Trained on VoxCeleb1 and Cn-Celeb1}
\begin{tabular}{ccccccccccc}
\toprule
\multirow{3}{*}{\textbf{Model}} & \multirow{3}{*}{\textbf{Score Norm}} & \multicolumn{4}{c}{\textbf{Training Dataset:VoxCeleb1}}                             & \textbf{} & \multicolumn{4}{c}{\textbf{Training Dataset:CN-Celeb1}}                             \\ \cline{3-6} \cline{8-11} 
                                &                                      & \multicolumn{2}{c}{\textbf{VoxCeleb1-O}} & \multicolumn{2}{c}{\textbf{CN-Celeb1-T}} & \textbf{} & \multicolumn{2}{c}{\textbf{VoxCeleb1-O}} & \multicolumn{2}{c}{\textbf{CN-Celeb1-T}} \\
                                &                                      & EER(\%)            & minDCF              & EER(\%)             & minDCF             &           & EER(\%)            & minDCF              & EER(\%)            & minDCF              \\ \hline
ECAPA-TDNN                      & -                                    & 2.23               & 0.157               & 11.97               & 0.457              &           & 8.22               & 0.463               & 10.43              & 0.453               \\
ResNet34                        & -                                    & 1.99               & 0.154               & 11.90               & 0.458              &           & 8.53               & 0.501               & 9.49               & 0.419               \\
Whisper-PMFA(ours)              & -                                    & 1.62               & 0.144               & 11.41               & 0.500              &           & 4.31               & 0.307               & 9.00               & 0.399               \\ \midrule
ECAPA-TDNN                      & AS-Norm                               & 2.00               & 0.144               & 12.12               & 0.405              &           & 7.73               & 0.432               & 10.10              & 0.403               \\
ResNet34                        & AS-Norm                               & 1.88               & 0.154               & 11.42               & \textbf{0.374}     &           & 7.87               & 0.454               & 9.27               & 0.385               \\
Whisper-PMFA(ours)              & AS-Norm                               & \textbf{1.42}      & \textbf{0.121}      & \textbf{11.24}      & 0.440              &           & \textbf{3.91}      & \textbf{0.270}      & \textbf{8.30}      & \textbf{0.358}      \\ \bottomrule
\end{tabular}
\vspace{-1em}
\end{table*}

\subsection{Implementation details}

To enhance the robustness of the systems, we applied three data augmentation techniques across all systems: additive noise augmentation from the MUSAN dataset\cite{snyder2015musan}, reverberation noise augmentation from the RIRs dataset\cite{ko2017study}, and speed perturbation\cite{yang2022data} with 0.9 and 1.1 times speed changes.

During the feature extraction phase for the two baseline models, we used Wespeaker's standard method. This process entails selecting random 2-second clips from each speech sample and extracting 80-dimensional FBank features from these segments. The window length was set to 25 milliseconds with a frameshift of 10 milliseconds, and no voice activity detection was performed.

For our proposed model, named Whisper-PMFA, we utilized an 80-channel log magnitude Mel spectrogram consistent with the training of Whisper. During the training phase, we initially froze the parameters of the Whisper model and fine-tuned the remaining parameters for 4 epochs. This strategy was employed to prevent the pre-trained model from being fine-tuned in the wrong direction due to the random initialization of other parts. Subsequently, we conducted an overall fine-tuning of the entire model.

All models were trained using the AAM-Softmax loss\cite{deng2019arcface,xiang2019margin}, with a margin of 0.2 and a scaling factor of 30. None of the models underwent large-margin fine-tuning during the training process.

\subsection{Evaluation}

We use cosine distance with AS-Norm for scoring. We report the system performance using two evaluation metrics: Equal Error Rate (EER) and Minimum Detection Cost Function (minDCF) with $P_{target}= 0.05$ and $C_{FA}= C_{Miss}= 1$.

\section{Evaluation results and analysis}
\subsection{Performance evaluation and analysis}

Experimental results are shown in Table 1. In the experiments, we conducted hierarchical fusion on layers 17-24, and the details regarding the selection of layers will be discussed in the next section.

As shown in Table 1, even though the two baseline models implemented within the Wespeaker framework achieved good results, our proposed model significantly outperforms the two baseline models on both the CN-Celeb1 and VoxCeleb1 datasets. On the VoxCeleb1 dataset, our model achieves an EER of 1.42\%, representing a reduction of 24.3\% compared to ResNet34 and a reduction of 29.0\% compared to ECAPA-TDNN. On the CN-Celeb1 dataset, our model achieves an EER of 8.30\%, representing a reduction of 10.4\% compared to ResNet34 and a reduction of 17.9\% compared to ECAPA-TDNN. MinDCF also shows corresponding improvements across all datasets.

The experimental results on multiple datasets demonstrate that although Whisper has not been optimized for speaker verification, our proposed method effectively utilizes the information learned and filtered from the pre-trained Whisper model. Whisper-PMFA leverages this information to capture distinctive speaker embeddings better, thereby significantly enhancing the effectiveness of speaker recognition tasks.

\subsection{Layer selection experiment}

The results of the layer selection experiment are presented in Table 2. We explored two different layer aggregation approaches: one using 8 layers and the other using 16 layers. In the 8-layer experiment, we divided the model into four parts: the front (layers 1-8), the front-middle (layers 9-16), the middle-back (layers 17-24), and the back (layers 25-32). In the 16-layer experiment, we selected adjacent parts based on the previous division, resulting in three groups of experiments: layers 1-16, layers 9-24, and layers 17-32~\footnote{Due to the constraint of GPU memory, a maximum of 16 layers is supported in our experiments}.

The experimental results for both the 8-layer and 16-layer experiments indicate that models utilizing 16 layers generally perform worse than those utilizing 8 layers. This suggests that increasing the number of layers does not necessarily improve model performance. Instead, the inclusion of excessively irrelevant information with an increased number of layers can hinder the model's pooling layers from effectively extracting relevant speaker-related cues. This ultimately leads to a degradation in performance.

Additionally, the experimental results indicate that compared to other parts, the mid-back part (layers 17-24) of the Whisper model contains more speaker-related cues. Fusion models based on these layers achieved the best performance.

\begin{table}[h]
\centering
\caption{Layer selection experiment result}
\begin{tabular}{ccc}
\toprule
\multirow{2}{*}{\textbf{Selected Layers}} & \multicolumn{2}{c}{\textbf{VoxCeleb1-O}}    \\
                                         & \textbf{EER(\%)}     & \textbf{minDCF}      \\ \midrule
1-8                                      & 3.93                 & 0.287                \\
9-16                                     & 1.66                 & 0.135                \\
17-24                                    & \textbf{1.42}        & \textbf{0.121}       \\
25-32                                    & 1.65                 & 0.148                \\ \midrule
1-16                                     & 2.07                 & 0.144                \\
9-24                                     & 1.74                 & 0.134                \\
17-32                                    & 2.03                 & 0.163                \\ \bottomrule
\end{tabular}
\end{table}
\vspace{-10pt}

\subsection{Cross-Language performance analysis}

The experiments in the second and third columns of the table demonstrate the cross-linguistic performance of our proposed model compared to the baselines. When trained on Chinese (CN-Celeb1) and tested on English (VoxCeleb1), as well as the reverse scenario, our proposed model consistently outperforms both baseline models. This phenomenon is particularly notable when training in Chinese and testing in English, where our model achieves performance improvements of nearly 50\% compared to the baselines. 
A critical factor contributing to this significant enhancement is utilizing the Whisper pre-trained model as the foundation of our proposed architecture. The Whisper model has been pre-trained on a diverse corpus that spans multiple languages, endowing it with robust cross-linguistic features. Our model, built upon this pre-trained base, inherits and optimizes these features for cross-lingual tasks. This advancement underscores the potential of pre-trained models in improving the adaptability and generalization of language processing systems across diverse linguistic contexts.

While our model achieves notable cross-lingual performance improvements in certain scenarios, its performance uplift is less marked when training in English (VoxCeleb1) and testing in Chinese (CN-Celeb1). This could be attributed to the complex nature of the CN-Celeb1 test set, which encompasses various domains such as drama, singing, and speeches. These diverse conditions likely impede the model's generalization in this context. Addressing this challenge will form the focus of our next research phase, aiming to enhance the model's adaptability and performance across varied linguistic domains.

\subsection{Comparison with results in the literature}

We compared our Whisper-PMFA with several state-of-the-art models, all models are trained on the VoxCeleb1 dataset. The experimental outcomes demonstrated that our proposed model achieved the highest performance among the models tested, this indicates the effectiveness of our method.


\vspace{-5pt}
\begin{table}[h]
\centering
\setlength{\tabcolsep}{5.2pt}
\caption{Comparison with published systems on VoxCeleb1}
\begin{tabular}{ccc}
\toprule
\multirow{2}{*}{\textbf{Model}} & \multicolumn{2}{c}{\textbf{VoxCeleb1-O}}    \\
                                         & \textbf{EER(\%)}     & \textbf{minDCF}      \\ \midrule
M-sv\cite{fan2020exploring}     & 3.61             & -               \\
Inter-layer Adapter WavLM\cite{sang2024efficient} & 2.58             & 0.187           \\
DROP-TDNN\cite{hong2023decomposition}& 2.15        & -               \\
HuBERT-Base ECAPA-TDNN\cite{chen2022large}& 1.86        & -               \\ \midrule
Whisper-PMFA(ours)              & \textbf{1.42}    & \textbf{0.121}  \\ \bottomrule
\end{tabular}
\end{table}
\vspace{-10pt}

\subsection{Investigation of more effective adaptation method}

To address the efficiency reduction caused by the Whisper model's size, we integrated LoRA\cite{hu2021lora} as an alternative to full fine-tuning. According to the results shown in Table 4, this method significantly reduced the trainable parameters by approximately 45 times with only a minimal increase in EER by 0.2\%. The Whisper-PMFA model, enhanced with LoRA, maintained high performance while achieving a parameter count comparable to other models, demonstrating a more effective adaptation method for large-scale models.

\vspace{-5pt}
\begin{table}[h]
\centering
\setlength{\tabcolsep}{4.8pt}
\caption{Comparison with the Integration of LoRA}
\begin{tabular}{cccc}
\toprule
\textbf{Model}         & \textbf{\# Params} & \textbf{EER(\%)} & \textbf{minDCF} \\ \hline
Whisper-PMFA           & 487.7M             & 1.42             & 0.121           \\
Whisper-PMFA(LoRA) & 10.9M              & 1.62             & 0.150           \\ \bottomrule
\end{tabular}
\end{table}
\vspace{-10pt}

\section{Conclusions}


In this paper, the Whisper-PMFA framework is proposed, which leverages the rich speech knowledge embedded in the pre-trained Whisper ASR model to achieve high-quality speaker embedding extraction through selective feature aggregation. Experimental results on the widely used VoxCeleb1 and CN-Celeb1 datasets show that Whisper-PMFA can achieve notably lower EERs than the competing models and high cross-linguistic robustness. In addition, the LoRA adaptation approach is also investigated as a trial adaptation method, achieving a significant reduction in the number of trainable model parameters while maintaining competitive performance.

\section{Acknowledgement}
Shuai Wang is supported by Internal Project of Shenzhen Research Institute of Big Data under grant No.J00220230014.

\bibliographystyle{IEEEtran}
\bibliography{mybib}

\end{document}